\documentclass{iaus}
\usepackage{graphicx}

\title[The role of quasars in galaxy formation] 
{The role of quasars in galaxy formation}

\author[D. Elbaz]   
{D. Elbaz}

\affiliation{Laboratoire AIM, CEA/DSM-CNRS-Universit\'e Paris Diderot, IRFU/Service d'Astrophysique, B\^at.709, CEA-Saclay, 91191 Gif-sur-Yvette C\'edex, France
              \\ email: {\tt delbaz@cea.fr} }

\pubyear{2009}
\volume{267}  
\pagerange{119--126}
\jname{Co-Evolution of Central Black Holes and Galaxies}
\editors{B.M.\ Peterson, R.S.\ Somerville, \& T.\ Storchi-Bergmann, eds.}
\begin{document}

\maketitle

\begin{abstract}
We discuss evidence that quasars, and more generally radio jets, may have played an active role in the formation stage of galaxies by inducing star formation, i.e. through positive feedback. This mechanism first proposed in the 70's has been considered as anecdotic until now, contrary to the opposite effect that is generally put forward, the quenching of star formation in massive galaxies to explain the galaxy bimodality, downsizing and the universal black hole mass over bulge stellar mass ratio. This suggestion is based on the recent discovery of an ultra-luminous infrared galaxies, i.e. an extreme starburst, which appears to be triggered by a radio jet from the QSO HE0450$-$2958 at $z$=0.2863, together with the finding in several systems of an offset between molecular gas and quasars, which may be explained by the positive feedback effect of radio jets on their local environment.
\end{abstract}

\firstsection 
\section{Introduction}
It has been realised over the past decade that the supermassive black hole (SMBH) at the centre of a galaxy bulge plays a major role in galaxy evolution. 
The process commonly considered is the negative feedback effect which may determine the final stellar mass of a galaxy so that it ends up with the locally observed universal bulge stellar mass over SMBH mass ratio (e.g. Cattaneo et al. 2009, Fabian 2009), M$_{\rm GAL}$/M$_{\rm BH}\sim$ 700 (500 in Marconi \& Hunt 2003, McLure \& Dunlop 2001 or Ferrarese et al. 2006, 700 in Kormendy \& Gebhardt 2001 and 830 in McLure \& Dunlop 2002).  Indeed the correlation that exists between the mass of the central SMBH of local galaxies and their bulge luminosity (Kormendy \& Richstone 1995, Magorrian et al. 1998), stellar mass or velocity dispersion (Gebhardt et al. 2000, Ferrarese \& Merritt 2000), suggests the existence of a physical mechanism connecting the activity of a galaxy nucleus and the bulding of its host galaxy stellar bulge. The most massive galaxies exhibiting the largest over-abundance of $\alpha$-elements with respect to the solar value, this means that they formed their stars the fastest (Thomas et al. 2005) and indeed a downsizing of the typical star forming galaxy has been observed in deep surveys, i.e. the comoving star formation rate activity of the Universe is dominated by more and more massive galaxies going back in lookback time. If negative feedback may provide a way to "kill" galaxies to reproduce the downsizing of galaxies, it does not necessarily explain why most stars in massive ellipticals formed rapidly. Hence positive feedback must have been at play early on shortly after the formation of the proto-galaxy. Major mergers have long been thought to be the best candidates to explain those rapid processes by first triggering a starburst then feeding the central SMBH and activating the QSO phase (Sanders et al. 1988), but this scenario has been subject to debate, in particular since it was found that AGN activity in the nearby Universe was not shown to correlate with galaxy pairs (Li et al. 2008). 

Since the masses of SMBH and of stars in galaxies are correlated, it is natural to consider the alternative possibility that both processes are related, with one possibly affecting the other. If radio jets are shown to be able to trigger star formation in galaxies, as suggested by early studies of distant galaxies but more or less abandoned since then, they may represent an interesting candidate process to understand the universal mass ratio since the most massive SMBH will have the largest accretion rates, hence also most powerful radio jets, possibly triggering more efficiently star formation. Because radio jets are difficult to identify, especially in the distant Universe, they have not yet been considered as a major actor in the process of galaxy formation. However, they must have been ubiquitous in the past to explain the ubiquity of SMBH in the center of local galaxies. Together with the following pieces of evidence, this suggests that their role may have been underestimated until now : (i) the discovery of a jet-induced ultra-luminous infrared galaxy (ULIRG), one of the strongest starbursts ever observed, in the system HE0450$-$2958, (ii) the presence of massive extended emission line regions (EELR) surrounding quasars and mostly radio galaxies, (iii) evidence that radio jets were more abundant in the past, (iv) the presence of large concentrations of molecular gas with an offset with respect to their neighboring QSO, possibly resulting from the impact of radio jets, (v) theoretical arguments suggesting that radio jets may trigger star formation efficiently.

Our aim here is not to firmly state that we are convinced that galaxy formation cannot be understood without accounting for the role of radio jets, since existing data and models are too sparse to either infer or reject such statement. However, in the near future, new observatories such as ALMA, eVLA, E-ELT or the JWST will provide crucial observations that should allow us to better address this issue, hence we wish here to bring back to the front this mechanism as a possible driver not only of star formation, but maybe even galaxy formation.

\section{The role of radio jets in star and galaxy formation} 
\label{SEC:jetinduced}
\begin{figure}
\begin{center}
\includegraphics[width=1.8in]{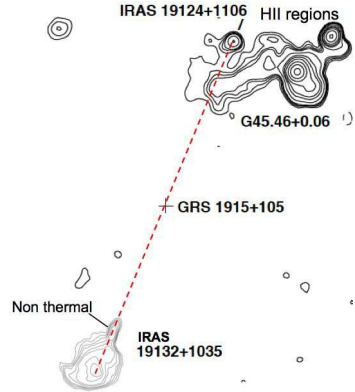} 
\caption{VLA mosaic at 20-cm of the surroundings of GRS 1915+105 (Rodriguez \& Mirabel 1998). The red dashed line illustrates the direction of the radio jets of the microquasar. }
\label{FIG:GRS1915}
\end{center}
\end{figure}

Evidence for jet-induced star formation has been found in various objects and environments, either far away from the host galaxy, such as in the lobe of radio jets (e.g. van Breugel et al. 1985), or inside the host galaxy, resulting in the so-called radio-optical alignment (e.g. McCarthy et al. 1987, Rees 1989). McCarthy et al. (1987) noticed that both the stellar continuum and size of the emission line region of 3CR radio galaxies at $z\ge$0.6 were highly elongated in parallel with their radio jets. This so-called radio--optical alignment was interpreted as evidence that the radio jets interact with the interstellar medium and stimulate large-scale star formation in the host galaxy (see also Rees 1989, Rejkuba et al. 2002, Oosterloo \& Morganti 2005).  Other mechanisms than jet-induced star formation have been suggested in the literature to explain the radio-optical alignment effect such as the scattering of light from the central AGN (e.g. Dey \& Spinrad 1996) or the nebular continuum emission from warm line-emitting regions (Dickson et al. 1995). However, many of these objects show clear evidence of star formation. This is, in particular, the case of 4C 41.17 ($z$=3.8), which rest-frame UV continuum emission is aligned with the radio axis of the galaxy, unpolarized and showing P Cygni-like features similar to those seen in star-forming galaxies (Dey et al. 1997). The most dramatic events, with star formation rates as high as 1000 M$_{\odot}$ yr$^{-1}$, are associated with very luminous radio galaxies at redshifts up to z $\sim$ 4 often located at the center of proto-clusters of galaxies (Dey et al. 1997, Bicknell et al. 2000, Zirm et al. 2005). Even in the closest radio galaxy Centaurus A, the reality of jet-induced star formation has been the center of a debate. Early on, Blanco et al. (1975) noticed the presence of extended gaseous filaments with bright knots exhibiting strong H$\alpha$ and UV emission or loose chains of blue compact objects aligned with the radio jets of the central active nucleus. The origin of these optical filaments has been subject to a long debate regarding the main source of their ionization until the young and massive stars inhabiting these optical knots were individually resolved showing that jet-induced star formation must indeed have taken place in this system (Rejkuba et al. 2002). If there is indeed a mechanism through which radio jets may induce star formation in quasars or radio galaxies, then this mechanism may also be at play inside their scaled-down version, i.e. microquasars. The relativistic jets of the microquasar GRS 1915$+$105, which show apparent superluminal motion, are aligned with two IRAS sources themselves associated with radio knots. One of the radio source was resolved with the VLA showing a non thermal extension pointing also in the direction of the microquasar on one side and the IRAS source on the other side (see Fig.~\ref{FIG:GRS1915}). Both IRAS sources appear to be H II regions ionized by late O or early B stars (Rodriguez \& Mirabel 1998).

Inversely, radio jets are known to be efficient in producing a negative feedback on the environment of the massive radio galaxies located at the center of the so-called "cooling flow" clusters (see Fabian 2009, Cattaneo 2009). There, even when its cooling time gets lower than 1 Gyr, the intracluster medium (ICM) temperature does not fall below a given threshold which implies that non gravitational heating is taking place. The presence of buoyant radio bubbles brings direct evidence that on these large scales, radio jets can prevent cooling from happening if they are long-lived. At the smaller scale of individual galaxies, there is no such strong direct evidence for negative feedback and the process must be different since radio jets are collimated, hence should escape galaxies without affecting the bulk of their ISM. For this reason, a second mode has been considered, called the radiative or wind mode (see discussion in Fabian 2009). Yet observationally, active nuclei are often found associated with star formation and both effects may not be exclusive in the sense that positive feedback may occur first at early stages in galaxy evolution followed by the quenching of star formation at a later stage. 

The discovery of a case of a ULIRG in which the intense star formation activity appears to be triggered by the impact of a radio jet (Elbaz et al. 2009 and Sect.~\ref{SEC:HE0450}) suggests that, at least in some cases, positive feedback from radio jets can be a powerful mechanism. This is a scaled up version of the prototypical example for jet-induced star formation, Minkowski's Object, and in both cases, no evidence has been found for the presence of stars older than the starburst which suggests that the process may be responsible for the triggering of the formation of a whole galaxy. The fact that at $z$$\sim$2, the comoving space density of radio galaxies with powerful jets was 1000 times higher than at the present epoch (Willott et al. 2001), suggests that the process taking place in these locally rare cases may have been an important actor in the process of galaxy formation and evolution. Relic traces of this activity may be seen in the presence of the extended emission line regions (EELRs) surrounding QSOs, and radio galaxies, with masses which may be as large as that of a massive galaxy, i.e. 10$^{10}$ M$_{\odot}$ (Fu \& Stockton 2008). Those gas reservoirs may represent the seeds for new galaxies if they are shocked by a radio jet as it may have happened in HE0450$-$2958 and such events may have been more common in the past. Indeed, hydrodynamical simulations of radiative shock-cloud interactions indicate that for moderate gas cloud densities ($>$ 1 cm$^{-3}$) such as those observed in EELRs, cooling processes can be highly efficient and result in more than 50\,\% of the initial cloud mass cooling to below 100 K, hence leading to star formation (Fragile et al. 2004). The mere existence of these EELRs itself is suggestive of the impact of radio jets on galaxies since their abundance ratios are similar to those observed in the host galaxies of their neighboring AGN, which Fu \& Stockton (2008) interpret as evidence that they were expelled out of the host galaxy by the radio jets.

Because they are collimated, radio jets may not affect very strongly the host galaxy of the active nucleus that produces them but they may produce an impact on the local environment. We present here two elements favoring this interpretation:

- the pair of ULIRGs in the HE0450$-$2958 system, discussed in the next section, provide evidence for a major event in the formation process of a galaxy related to radio jets (see Sect.~\ref{SEC:HE0450}).

- several AGNs have been found with neighboring molecular gas clouds, which may be the result of the impact of a radio jet on its environment (see Sect.~\ref{SEC:CO}).

\section{The case of HE0450$-$2958} 
\label{SEC:HE0450}
HE0450$-$2958 is a nearby luminous ($M_V$$=$$-25.8$) radio quiet quasar located at a redshift of $z$=0.2863 (Canalizo \& Stockton 2001) and presently the only known quasar for which no sign of a host galaxy has been found (Magain et al. 2005). This object has been the center of a debate and various scenarios have been proposed to explain the absence of detection of a host galaxy. Since dust extinction could potentially explain it, we imaged HE0450$-$2658 with VISIR, the Very Large Telescope Imager and Spectrometer in the Infrared (VISIR) at the ESO-VLT. Indeed, a bright source was detected by IRAS at the approximate location of HE0450$-$2958 within the position error bar of 5 arcmin which, at the luminosity distance of the QSO ($z$=0.2863), would translate into the total infrared luminosity of an ultra-luminous infrared galaxy (ULIRG, L$_{\rm IR}$=L(8--1000\,$\mu$m)$\geq$10$^{12}$ L$_{\odot}$, de Grijp et al. 1987, Low et al. 1988). In a first analysis, a unique mid-IR source was detected with VISIR associated with the QSO, but combined with HST--NICMOS imaging at 1.6\,$\mu$m, it was not possible to disentangle a local dust torus from a whole dust-obscured galaxy (Jahnke et al. 2009). However, a re-analysis of the VISIR data led to the detection of a second mid-IR source in the field, associated with the 7 kpc distant companion galaxy to the QSO (Elbaz et al. 2009). 
\begin{figure}[h]
\begin{center}
 \includegraphics[width=2.2in]{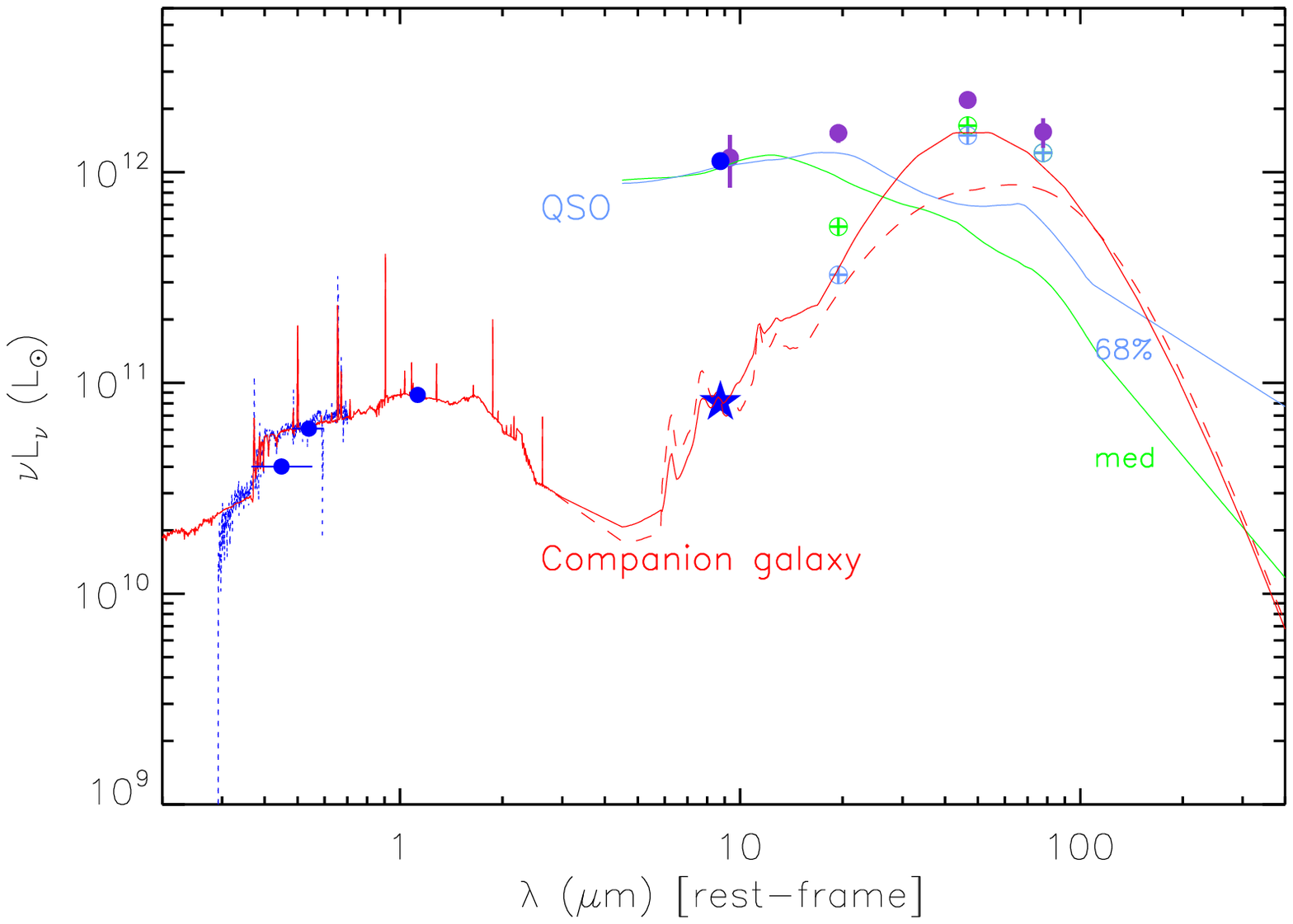} 
 \includegraphics[width=1.4in]{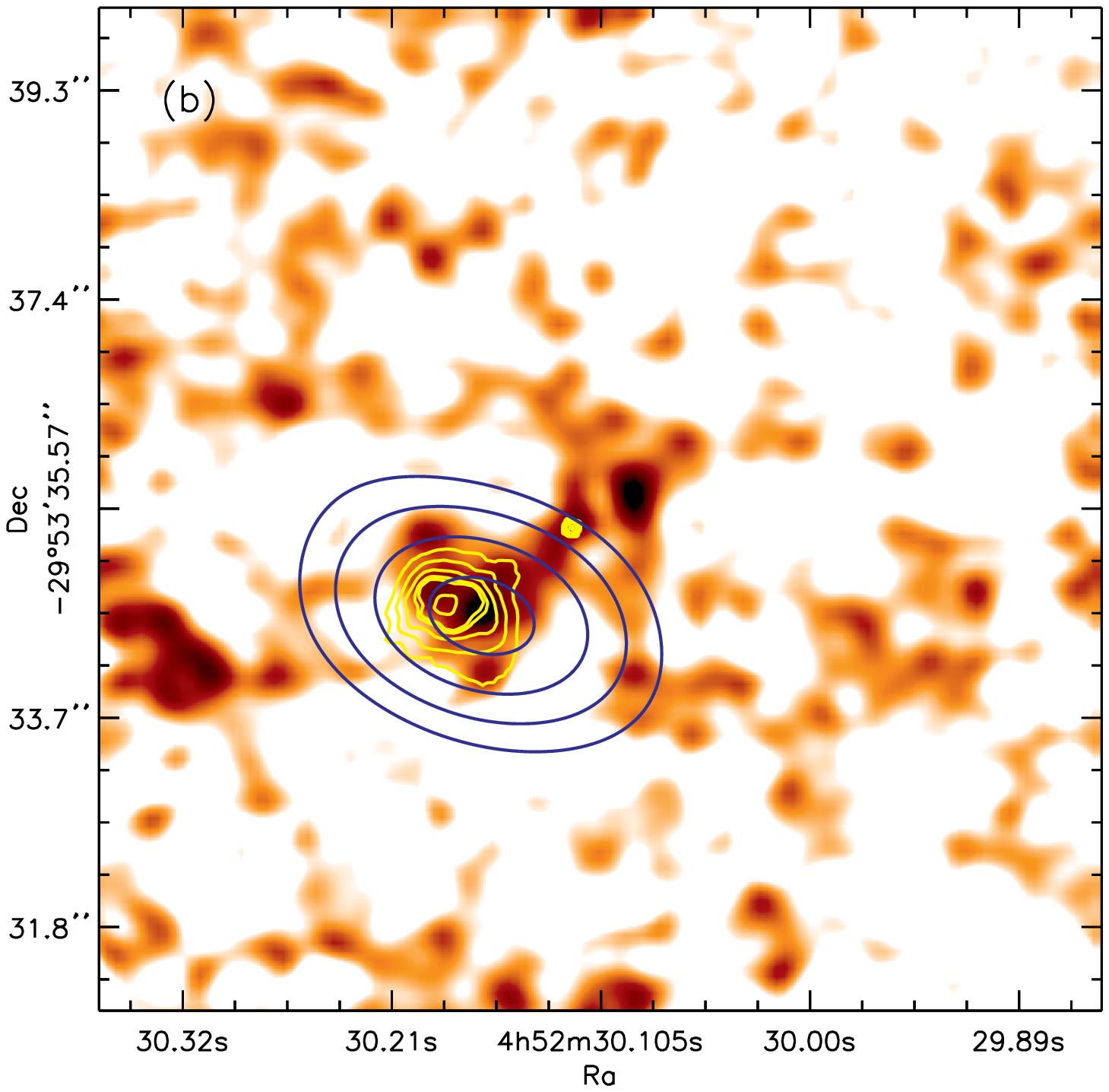} 
 \includegraphics[width=1.4in]{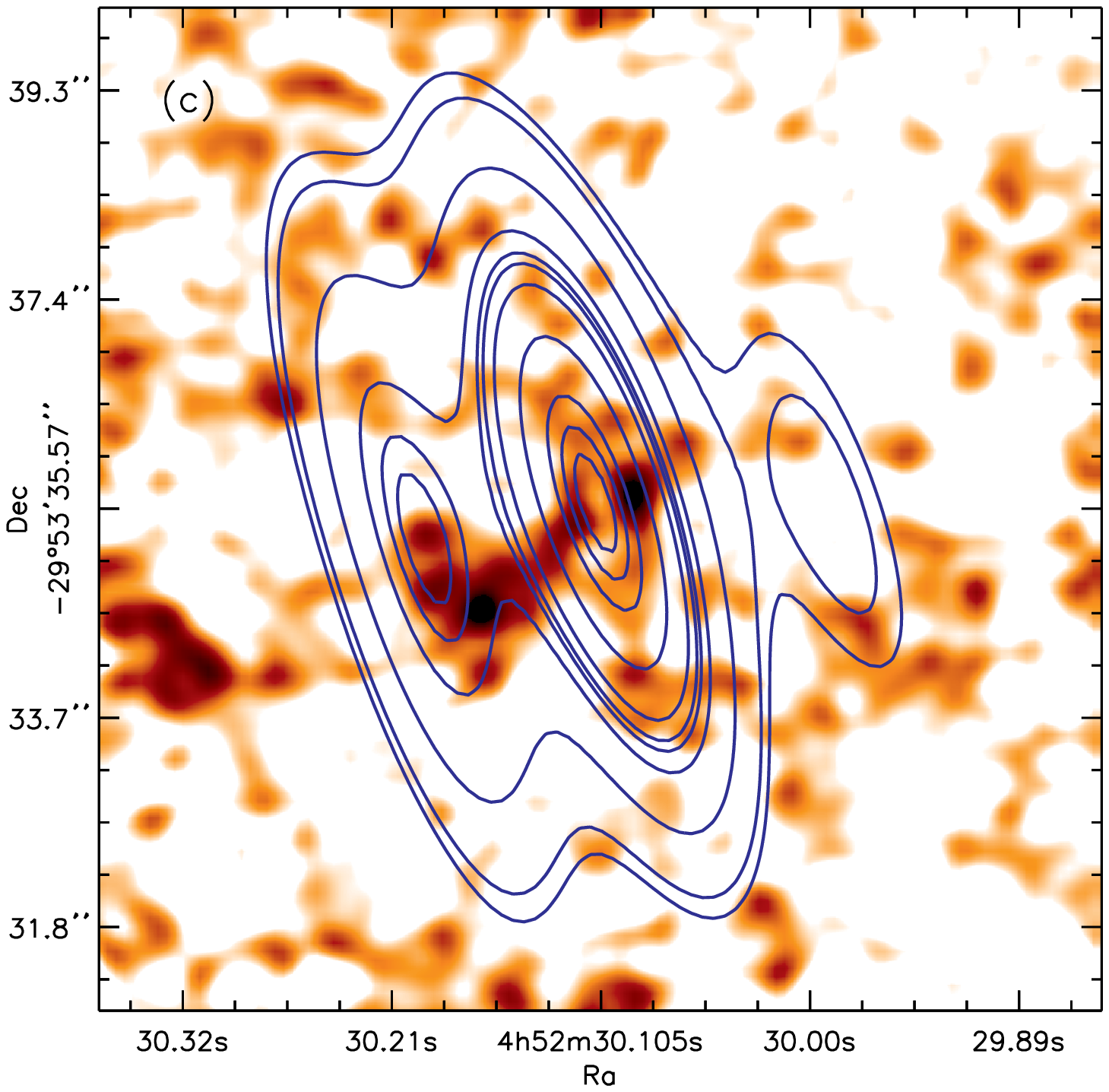} 
\caption{\textbf{\textit{(a)}} Full spectrum of the companion galaxy of HE0450$-$2958 from 0.3 to 400\,$\mu$m in the rest-frame (red line). The plain green and blue lines represent the median and 68 percentile reddest SEDs from the Atlas of 47 quasar SEDs of Elvis et al. (1994), normalized to the VISIR measurement at 11.3\,$\mu$m (observed; 8.9\,$\mu$m rest-frame) associated to the QSO (filled blue circle). Filled blue star for the companion galaxy. \textbf{\textit{(b)}} VISIR image with contours from NICMOS-F160W at 1.6\,$\mu$m (in yellow, 1.2\,$\mu$m in the rest-frame, Jahnke et al. 2009) and ATCA CO 1--0 (dark blue lines, from Papadopoulos et al. 2008). \textbf{\textit{(c)}} ATCA 6208 MHz radio continuum (dark blue lines, from Feain et al. 2007) emission associated with the radio jets overlayed on the VISIR image.}
   \label{FIG:im}
\end{center}
\end{figure}
The VISIR image provides the first direct evidence that the IRAS "source" is in fact made of two well separated objects, the QSO and its companion galaxy. The infrared SED of the "system" (Fig.~\ref{FIG:im}a) peaks around $\sim$50\,$\mu$m (rest-frame) which indicates that the bulk of the far infrared light arises from star formation (the QSO and companion galaxy are both contained in the single IRAS measurement) and not from and active nucleus that would drop beyond $\sim$20\,$\mu$m (Netzer et al. 2007, minimum temperature of $\sim$200 K typically). The IR SED of the companion galaxy, which produces the bulk of the far-IR light, is very well fitted with a standard star forming template SED (red plain line, from the library of Chary \& Elbaz 2001). The total infrared luminosity of the companion galaxy is L$_{\rm IR}$$\sim$2$\times$10$^{12}$ L$_{\odot}$, which corresponds to a SFR$\sim$340 M$_{\odot}$yr$^{-1}$ (conversion factor of Kennicutt 1998). Hence HE0450$-$2958 appears to be a composite system made of a pair of ULIRGs (both galaxies have similar L$_{\rm IR}$), where both mechanisms, star formation and QSO activity, are spatially separated by 7 kpc in two distinct sites. The fact that the starburst is associated with the companion galaxy is reinforced by the finding that the molecular gas traced by the CO molecule appears to avoid the QSO but to peak at the location of the companion galaxy (dark blue contours in Fig.~\ref{FIG:im}b) close to the peak mid-infrared emission measured with VISIR (Papadopoulos et al. 2008). This both suggests that there is a large amount of dust in this galaxy and that there is a large gas reservoir (M(H$_2$)$\simeq$2.3$\times$10$^{10}$ M$_{\odot}$) to fuel an intense star formation event. No evidence for the presence of an old stellar population is found in the VLT-FORS optical spectrum of the companion, although the radiation of such stars could be diluted in that of the youngest population. The age of the dominant stellar population is about 40--200 Myr suggesting that the galaxy was recently born. The stellar mass of the galaxy (M$_{\star}$$\simeq$[5--6]$\times$10$^{10}$ M$_{\odot}$) is consistent with this timescale for the SFR derived from its IR luminosity.

\begin{figure}[h]
\begin{center}
 \includegraphics[width=1.98in]{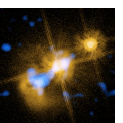} 
\includegraphics[width=3.02in]{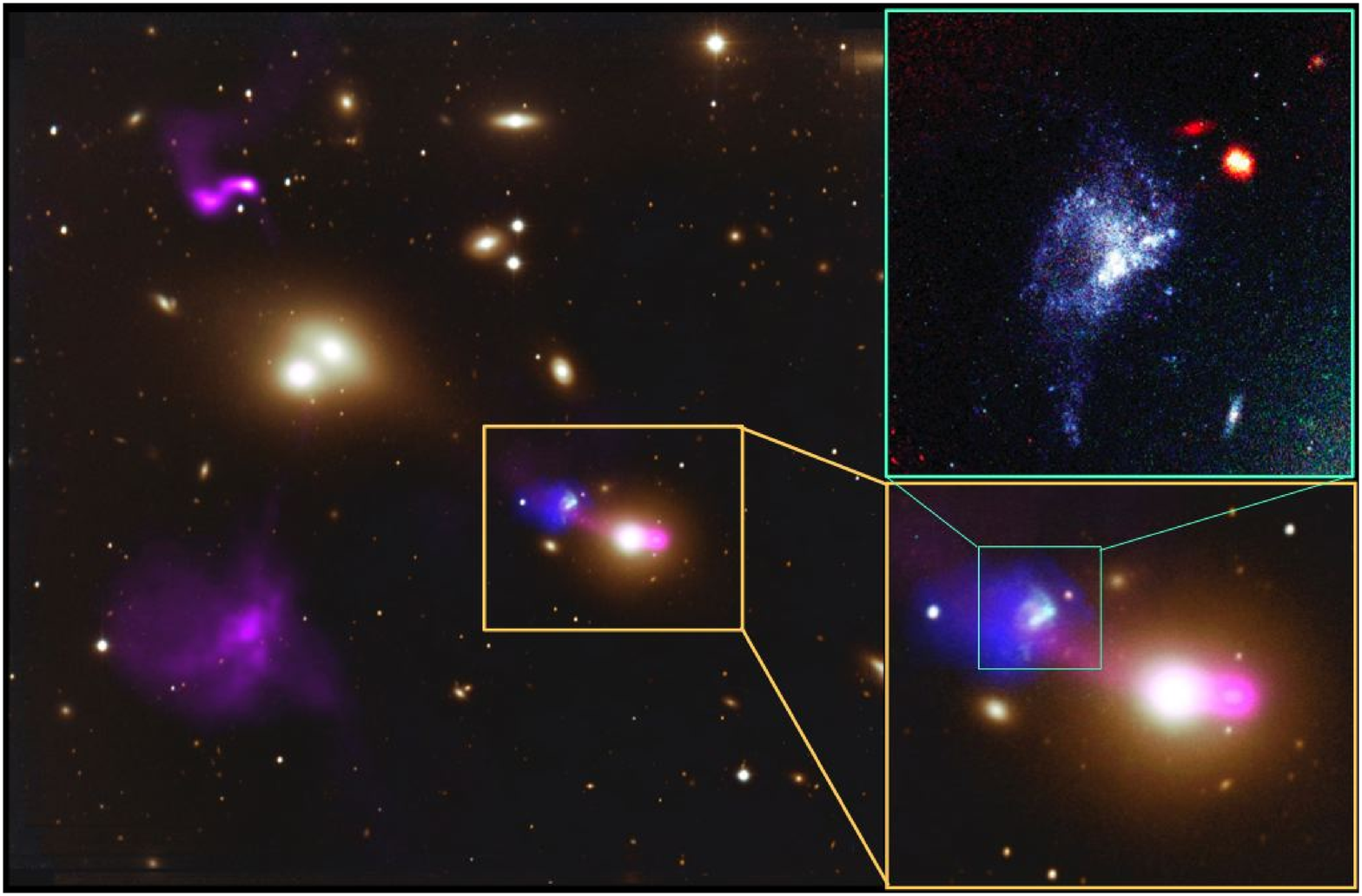}
\caption{\textbf{\textit{(left)}} composite image of the HE0450$-$2958 system with VISIR 11.3\,$\mu$m image (blue) and HST in the R-band (yellow). \textbf{\textit{(right)}} Composite image showing the details of Minkowski's Object (Croft et al. 2006). Optical image of the field of Abell 194, overlaid with radio continuum ( purple), HI data (dark blue), and H$\alpha$ data (light blue).}
   \label{FIG:jets}
\end{center}
\end{figure}

The companion galaxy coincides with one of the two radio lobes on both sides of the QSO HE0450$-$2958 (Fig.~\ref{FIG:im}c), suggesting that it is hit by a radio jet emitted by the QSO. Evidence for shock excitation of the ISM in the companion galaxy, as traced by an high [NII]/H$\alpha$ emission line ratio (Letawe et al. 2008b), confirms that the association of the jet and the galaxy is physical and not due to a projection effect.
The VISIR image itself shows the presence of a bridge of mid-IR emission between the QSO and the companion Fig.~\ref{FIG:jets}-left.
. Altogether, this suggests that it is the radio jet that is triggering the starburst in the companion. As a result, HE0450$-$2958 is comparable to Minkowski's Object, a proto-typical case for jet-induced star formation (van Breugel et al. 1985). Minkowski's Object is a newly formed galaxy, with a stellar population age of only 7.5 Myr and a stellar mass of 1.9$\times$10$^7$ M$_{\odot}$ (Croft et al. 2006), aligned with the radio jet of NGC 541, an FR I galaxy (Fanaroff \& Riley 1974) located in the cluster Abell 194. The companion galaxy of HE0450$-$2958 might be a scaled-up version of Minkowski's Object, i.e. a massive galaxy whose formation was jet-induced. 

Evidence for the presence of large amounts of matter in the surrounding of radio quasars, in the form of massive gas clouds exhibiting strong emission lines with high excitation levels, has been gathered since the early 60s. These extended emission lines regions (EELRs, see Stockton \& MacKenty 1983) can reach masses of several 10$^{10}$ M$_{\odot}$ (Fu \& Stockton 2008, Stockton et al. 2007). Spectro-imaging and narrow-band filter imaging in the fields of radio quasars demonstrated that EELRS were made of clouds of gas, whose strong emission lines were excited by the hard radiation of the quasar. Fu \& Stockton (2008) found that low metallicity quasar host galaxies happened to be surrounded by EELRs with similar abundance ratios, which they interpreted as evidence that the EELRs had been expelled from the host galaxy itself. The double-lobed morphologies of extragalactic radio sources shows that the relativistic jets not only couple strongly with the ISM of the host galaxies, but are capable of projecting their power on the scales of galaxy haloes and clusters of galaxies (up to $\sim$1 Mpc, Tadhunter 2007). 

Several EELRs were found in the surroundings of HE0450$-$2958 following the axis of the radio jets. They have been detected up to 30 kpc away from the QSO (Letawe et al. 2008a). The N-W blob of gas, photoionized by the radiation of HE0450$-$2958, is itself located in the direction of the radio jet opposite to the one pointing at the companion galaxy. The thermodynamic state of EELRs is uncertain since the densities one can infer from their optical spectra are of the order of n$_{\rm e}$=100--300 cm$^{-3}$ and T$_{\rm e}$=10,000--15,000 K (Fu \& Stockton 2008), which imply that they should be gravitationally unstable and have already started forming stars. Hence Fu \& Stockton (2008) modeled these regions using two components, a low density component, where most of the mass is locked at densities of order n$_{\rm e}\sim$1 cm$^{-3}$, and a higher density component responsible for the observed emission lines. Hydrodynamical simulations of radiative shock-cloud interactions indicate that for moderate gas cloud densities ($>$ 1 cm$^{-3}$), cooling processes can be highly efficient and result in more than 50\,\% of the initial cloud mass cooling to below 100 K (Fragile et al. 2004). Hence, the companion galaxy of HE0450$-$2958 could have formed from a seed EELR that was hit by the radio jet. This process could explain why radio jets do not always produce the same effect.

\section{A test of jet-induced galaxy formation for future instrumentation: offset molecular gas} 
\label{SEC:CO}
Here we search for other cases such as HE0450$-$2958, where a large mass of molecular gas is found offset with respect to the QSO and associated with newly formed stars which formation could be induced by a radio jet. Such an offset between molecular gas and QSOs may be a common feature in distant radio sources as suggested by Klamer et al. (2004) who produced a systematic search in $z$$>$3 CO emitters for an AGN offset with respect to the CO concentrations and for a connection with radio jets. Out of the 12 $z>$ 3 CO emitters that they studied, six showed evidence of jets aligned with either the CO or dust, five have radio luminosities above 10$^{27}$ W Hz$^{-1}$ and are clearly AGNs, and a further four have radio luminosities above 10$^{25}$ W Hz$^{-1}$ indicating either extreme starbursts or possible AGNs. In the following, we discuss three proto-typical cases classified with increasing redshift from $z$$=$2.6 to 6.4 that may be compared with HE0450$-$2958.

At $z$=1.574, 3C18 is radio loud quasar with recently formed radio jets, as inferred from their small physical size. A large mass of molecular gas ($M_{\rm H_2}$$\sim$(3.0$\pm$0.6)$\times$10$^{10}$ M$_{\odot}$), inferred from CO 2--1 emission, is found associated to the quasar with a positional offset of $\sim$20 kpc (Willott et al. 2007).

At $z$= 2.6, TXS0828$+$193 is a radio galaxy with a neighboring CO gas concentration of $\sim$1.4$\times$10$^{10}$ M$_{\odot}$ located 80 kpc away and shows no evidence for an underlying presence of stars or galaxy (Nesvadba et al. 2009). An upper limit of 0.1 mJy at 24\,$\mu$m, from the MIPS camera onboard Spitzer, was obtained by Nesvadba et al. (2009), who concluded from this limit that no major starburst with more than several hundred solar masses per year could be taking place associated with this CO concentration. We used the library of SED templates of local galaxies from Chary \& Elbaz (2001) to convert this mid infrared measurement into an upper limit for the total IR luminosity at this position of L$_{\rm IR}^{\rm max}$$\sim$7.8$\times$10$^{12}$ L$_{\odot}$, which would translate to a maximum SFR of $\sim$1340 M$_{\odot}$ yr$^{-1}$ using the Kennicutt (1998) conversion factor for a Salpeter IMF. Hence there is still room for a large amount of star formation in this object that may be in a similar stage than HE0450$-$2958 but at a much larger distance. 

At $z$= 4.695, BR 1202$-$0725 is the first high-redshift quasar for which large amounts of molecular gas was detected (Omont et al. 1996). 
The CO map presents two well separated emission peaks which coincide with radio continuum emission interpreted as evidence for the presence of radio jets by Carilli et al. (2002). The radio jets themselves would be too faint to be detected at the sensitivity level of the radio image but marginal evidence for variability suggest that the radio emission is not due to star formation. Omont et al. (1996) suggested that the double CO emission might be due to gravitational lensing, but no optical counterpart of the QSO is found associated with the second source and the CO (2--1) line profiles are different for the two components (Carilli et al. 2002). Klamer et al. (2004) suggested that stars may have formed along the radio jets, providing both the metals and dust for cooling and "conventional" star formation to take place afterwards.

In the case of HE0450$-$2958, the mid infrared emission is spatially associated with the offset CO concentration suggesting that similar systems might be found not only through CO imaging but also infrared imaging. SDSS160705+533558, located at a redshift of $z$=3.65, might be a distant analog of HE0450$-$2958 in that respect since it also presents a positional offset between the maximum sub-millimeter (from the submillimeter arra, SMA) and optical emission (Clements et al. 2009). 

A key test for the role of radio jets in the formation of galaxies will be the detection of either molecular gas or mid infrared emission with a positional offset with respect to their neighboring QSO. This test will be fulfilled with the advent of ALMA, the JWST or project instruments such as METIS (mid infrared E-ELT Imager and Spectrograph) for the ELT (extremely large telescope).

\end{document}